\documentclass[prl,twocolumn,amsmath,amsfonts,floatfix,showpacs,letterpaper]{revtex4}

\usepackage{graphicx}
\usepackage{ifpdf}

\DeclareMathOperator{\Tr}{Tr}

\newcommand{\beq}[1]{\begin{equation}\label{#1}}
\newcommand{\eeq}{\end{equation}}
\newcommand{\refeq}[1]{Eq.~(\ref{#1})}

\newcommand{\beqm}[1]{\begin{multline}\label{#1}}
\newcommand{\eeqm}{\end{multline}}

\newcommand{\punc}[1]{\,{\text{#1}}}
\newcommand{\sub}[1]{_{\mathrm{#1}}}

\newcommand{\kagome}{kagom\'e}

\newcommand{\Z}{{\mathcal{Z}}}
\newcommand{\T}{{\mathcal{T}}}
\newcommand{\Ham}{{\mathcal{H}}}

\newcommand{\Lag}{{\mathcal{L}}}

\newcommand{\clH}{\mathcal{E}}
\newcommand{\quH}{\mathcal{H}}

\newcommand{\Rot}{\mathbb{R}}

\newcommand{\nd}{^{\phantom{\dagger}}}

\newcommand{\Av}{{\mathbf{A}}}
\newcommand{\del}{\boldsymbol{\nabla}}
\newcommand{\phiv}{\boldsymbol{\varphi}}
\newcommand{\sigv}{\boldsymbol{\sigma}}
\newcommand{\mv}{\mathbf{m}}

\newcommand{\ev}{\mathbf{e}}
\newcommand{\dv}{\boldsymbol{\delta}}

\newcommand{\rv}{{\mathbf{r}}}

\newcommand{\zerov}{{\boldsymbol{0}}}

\newcommand{\ee}{\mathrm{e}}
\newcommand{\ii}{\mathrm{i}}

\usepackage{color}

\newcommand{\putinscaledfigure}[1]{\begin{center}\includegraphics[width=0.9\columnwidth]{#1}\end{center}}

\begin{document}

\title{$\mathrm{SU}(2)$-invariant continuum theory for an unconventional phase transition in a three-dimensional classical dimer model}

\author{Stephen Powell}
\author{J.\ T.\ Chalker}
\affiliation{Theoretical Physics, Oxford University, 1 Keble Road, Oxford, OX1 3NP, United Kingdom}

\begin{abstract}
We derive a continuum theory for the phase transition in a classical dimer model on the cubic lattice, observed in recent Monte Carlo simulations. Our derivation relies on the mapping from a three-dimensional classical problem to a two-dimensional quantum problem, by which the dimer model is related to a model of hard-core bosons on the \kagome\ lattice. The dimer-ordering transition becomes a superfluid--Mott insulator quantum phase transition at fractional filling, described by an $\mathrm{SU}(2)$-invariant continuum theory.
\end{abstract}

\pacs{
64.60.Bd, 
64.70.Tg, 
75.10.Hk  
}

\maketitle

The standard model of symmetry-breaking phase transitions, both classical and quantum, is the Landau-Ginzburg-Wilson (LGW) theory \cite{Landau}, where the critical properties are described by a continuum theory written in terms of the order parameter of the transition. It has recently been argued, however, that in certain two-dimensional quantum systems, continuous phase transitions are possible between symmetry-breaking states with apparently unrelated order parameters, in conflict with the LGW paradigm \cite{Senthil, Balents}.

Another class of non-LGW transitions occurs in classical systems with constraints that prevent a fully disordered state \cite{Alet}. As the temperature is raised, these systems instead enter a `Coulomb phase', where correlation functions have power-law forms and strong directional dependence. A na\"\i ve application of the LGW theory fails to capture these correlations and so cannot describe a transition between an ordered phase and a Coulomb phase \cite{Bergman}.

Recent numerical work \cite{Alet,Misguich} indicates that such a transition exists in a classical dimer model on the cubic lattice. This model describes the statistics of close-packed `dimers', objects that occupy two neighbouring sites of the lattice, with every site of the cubic lattice covered by precisely one dimer. There are many configurations that obey this constraint, and if all are given equal weight, the system displays a Coulomb phase \cite{Huse}. If instead they are given Boltzmann weights that favour parallel dimers, the system orders at low temperatures; the ordered phase is a six-fold degenerate crystal, breaking the lattice symmetry.

In this Letter, we outline two steps that lead to a continuum description of this classical dimer transition. Our first step uses the standard mapping between classical statistical mechanics in 3D and quantum mechanics in 2D, and so provides a bridge between the two classes of proposed non-LGW transitions. In the second step, we show that long-wavelength properties at the resulting 2D quantum transition are described by the $\mathrm{SU}(2)$-symmetric noncompact $CP^1$ (NC$CP^1$) model. This conclusion is consistent with earlier suggestions, on the basis of results for several three-dimensional (3D) quantum models at finite temperature \cite{MotrunichSenthil,Bergman}, that the classical dimer transition should be described by a gauge theory coupled to multiple matter fields.

Our approach is to identify the $[111]$ direction as imaginary time and map to a model of hard-core bosons on the \kagome\ lattice. A related mapping has previously been applied to another classical transition that is believed to lie outside the LGW paradigm, the Kasteleyn transition of spin ice in a $[100]$ magnetic field \cite{SpinIce}. As in that case, the Coulomb phase of the classical problem maps onto a superfluid phase for the quantum bosons, and the long-range power-law form for the correlation functions is reproduced by the fluctuations of the Goldstone (phase) mode. The dimer crystal becomes a bosonic Mott insulator, with a charge-density-wave order that breaks the symmetry of the \kagome\ lattice. There is a six-fold degeneracy in the ordering pattern, corresponding to the degeneracy of the crystalline phase.

In contrast to the case of spin ice \cite{SpinIce}, the quantum phase transition that results is still not amenable to a straightforward application of the LGW theory, and is instead an example of the class of non-LGW quantum transitions described above. The transition is between a bosonic superfluid and a Mott insulator at fractional filling, and is of the type considered by Balents et al.~\cite{Balents}. They showed that, using a dual representation in terms of vortices, the continuum behaviour can be described by a gauge theory whose form is strongly constrained by the symmetries of the original boson problem. In this dual model, the condensed phase of bosons becomes the noncondensed phase of the vortices, where the fluctuations of the photon mode of the gauge theory reproduce the superfluid correlations. The condensation of the vortex fields leads, by the Anderson-Higgs mechanism, to a gap for the photon and short-ranged correlation functions.

In our case, the continuum theory is an $\mathrm{SU}(2)$-invariant gauge theory of a two-component complex vector $\phiv$ minimally coupled to a noncompact $\mathrm{U}(1)$ gauge field $\Av$,
\beq{Lag0}
\Lag _0 = |(\del - \ii\Av)\phiv|^2 + s|\phiv|^2 + u(|\phiv|^2)^2 + \kappa |\del \times \Av |^2\punc{.}
\eeq
This can be rewritten in the usual form for NC$CP^1$ by replacing the terms $s$ and $u$ by a hard constraint $|\phiv| = 1$. Although the method picks out one direction as imaginary time, space-time isotropy is restored in the final form of the theory, a 3D classical continuum model ($\del$ denotes the three-dimensional derivative operator).


High-precision Monte Carlo data \cite{Alet} suggest that the dimer model has either a single continuous transition as the temperature is varied, or a very weak first-order transition, with a large but finite correlation length. In both cases, we expect there to be a well-defined continuum regime with correlation length much longer than the lattice spacing, described by \refeq{Lag0}. It remains controversial whether this continuum theory itself exhibits a continuous transition \cite{Motrunich1,FJJiang,Motrunich2,Kuklov}.

An outstanding puzzle noted in Refs.~\cite{Alet} and \cite{Misguich}, which we are unable to resolve, is that the critical exponent values extracted from simulations of the dimer model \cite{Alet} differ from those obtained in direct simulations of the NC$CP^1$ model \cite{Motrunich1,Motrunich2,Kuklov} and related quantum spin models \cite{Sandvik,Melko}. It has been suggested \cite{Alet,Misguich} that the discrepancy may be due to the proximity of a tricritical point in the phase diagram of a generalized dimer model; further numerical work is required to clarify this point as well as the nature of the transition.


The configurations of the classical model are close-packed arrangements of hard-core dimers, described by the variables $d_\mu (\rv) \in \{0,1\}$, giving the occupation number of the link between the cubic-lattice sites $\rv$ and $\rv + \dv _\mu$, where $\dv _\mu$ is a unit vector ($\mu \in \{x,y,z\}$). The close-packing constraint can be expressed as $\sum _{\mu} \left[ d_\mu(\rv) + d_\mu(\rv - \dv _\mu) \right] = 1$, for all $\rv$. The energy of a configuration is $\clH = - n_\parallel$, where $n_\parallel$ is the number of plaquettes (of any orientation) with parallel dimers, though our continuum theory applies equally to other potentials that also favour columnar crystalline order.

At zero temperature, $T = 0$, the dimers maximize $n_\parallel$ by selecting one of the six degenerate columnar ordering patterns, distinguished by the staggered order parameter,
\beq{Magnetization}
m_\mu(\rv) = \frac{1}{2}(-1)^{r_\mu} [d_\mu(\rv) - d_\mu(\rv - \dv _\mu)]\punc{.}
\eeq
The six ground states have $\mv(\rv) \in \{ {\pm \dv _x},{\pm \dv _y},{\pm \dv _z} \}$, for all $\rv$. For $T = \infty$, $\langle \mv \rangle = \zerov$, and the dimer--dimer correlation function has the 3D dipole form \cite{Huse}
\beq{CoulombCorrelators}
\langle d_\mu(\rv) d_\nu(\zerov) \rangle \sim \eta _{\rv} \frac{3 r_\mu r_\nu - |\rv|^2 \delta_{\mu\nu}}{|\rv|^5}\punc{,}
\eeq
where $\eta_\rv = (-1)^{\sum_{\mu}\!r_\mu}$ is $\pm 1$ on the two sublattices.

Monte Carlo simulations show a transition, apparently continuous, from an ordered phase to a Coulomb phase as $T$ is increased, at a critical temperature of $T\sub{c} \simeq 1.675$ \cite{Alet}. For $T < T\sub{c}$, the lattice symmetry is broken and $\langle\mv\rangle$ takes on a nonzero value (oriented along one of the cubic unit vectors $\pm \dv _\mu$). For $T > T\sub{c}$, the order parameter vanishes and the correlation functions have a dipolar form.

\begin{figure}
\putinscaledfigure{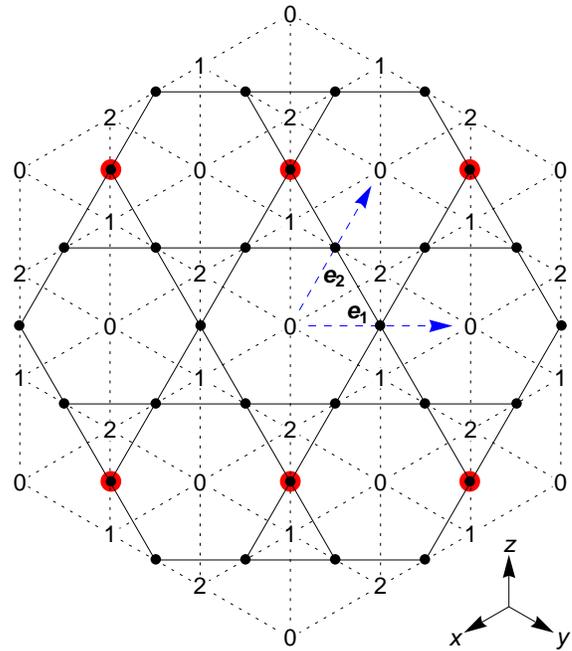}
\caption{\label{Cubic111Projection}The projection of the cubic lattice onto a $(111)$ plane, with the \kagome\ lattice superimposed. The cubic sites are shown by the numbers $0$, $1$ and $2$, giving $\tau \bmod 3$ (see main text). Points show the sites of the \kagome\ lattice, situated at the centres of the cubic bonds between sites with $\tau \bmod 3 = 1$ and $2$; they therefore lie in planes with $\tau \bmod 3 = \frac{3}{2}$. The larger red circles shown superimposed on some of the \kagome\ sites illustrate the occupied sites in one of the six degenerate ordering patterns, corresponding to the six ordered states of the cubic dimers. The elementary unit vectors of the \kagome\ lattice, $\ev _1$ and $\ev _2$, are shown with dashed blue arrows. The coordinate axes shown in the bottom-right of the figure are the projection of the cubic $x$-, $y$- and $z$-axes onto the $(111)$ plane.}
\end{figure}

We now outline a derivation of the continuum theory for this transition, based on a mapping to quantum bosons in 2D. Our first step is to define the imaginary time $\tau = \sum_\mu r_\mu$, given by the projection of the 3D position onto the $[111]$ direction. We then map the configurations of a given $(111)$ plane of the classical problem to the basis states in a quantum Hilbert space, by simply identifying the presence (or absence) of a dimer with the presence (absence) of a boson. As illustrated in Figure~\ref{Cubic111Projection}, the midpoints of the cubic bonds satisfying $\tau \bmod 3 = \frac{3}{2}$ form a set of \kagome\ planes stacked in the $[111]$ direction, and we use this set as the space-time lattice for the quantum problem.

For this mapping to make sense, we require conservation of particle number. The close-packing constraint implies that
\beq{ClosePacking2}
\sum _{\rv \in \tau} \sum _\mu d_\mu(\rv) = \sum _{\rv \in (\tau - 1)} \left [ 1 - \sum _\mu d_\mu(\rv) \right]\punc{,}
\eeq
where $\rv \in \tau$ indicates a sum over all cubic sites in imaginary-time slice $\tau$. This implies that if the \kagome\ plane at $\tau = \frac{3}{2}$ has $n$ bosons and a total of $N$ sites, then the plane at $\tau = \frac{3}{2} + 3$ will have $\frac{N}{3}-n$ bosons. We must therefore choose an imaginary-time step $\delta\tau$ that is even in order to give particle conservation; the smallest choice is $\delta\tau = 6$ and so we take only the bonds whose midpoints are at $\tau \bmod 6 = \frac{3}{2}$ to define the quantum problem.

The six distinct ground states of the dimers map to six ordered states of the \kagome\ bosons, and the example with $\mv = -\dv _z$ is shown in Figure~\ref{Cubic111Projection}. In each ordered state, half of the sites on a given sublattice are occupied, so the density (bosons per site) is $f = \frac{1}{6}$. The low-temperature phase of the classical model maps to the ordered phase of the bosons, while in the high-temperature phase the bosons condense and there is no spatial order, corresponding to the Coulomb phase with $\langle \mv \rangle = \zerov$.

The basis states for the quantum Hilbert space are given by the occupation-number states of hard-core bosons on the \kagome\ lattice. The hard-core nature of the dimers further implies that there is hard-core nearest-neighbour repulsion, and each triangle of the \kagome\ lattice (of either orientation) can be occupied by at most one boson. The partition function for the classical problem is $\Z = \Tr \T^{L/\delta\tau}$, where $L$ is the length of the system in the $[111]$ direction and periodic boundary conditions are assumed in this direction. The transfer matrix $\T$ has rows and columns labelled by the configurations of two planes separated by $\delta\tau = 6$, and its elements give the statistical weights for these configurations, summed over all possible configurations of the intermediate planes. The effective quantum Hamiltonian $\quH$ is defined by $\T = \ee^{-\quH \delta\tau}$, so that $\Z$ is given by the quantum partition function at inverse temperature $\beta \propto L$. The classical thermodynamic limit is therefore given by the quantum zero-temperature limit, $\beta \rightarrow \infty$.

It is in principle possible to find $\T$, and hence $\quH$, exactly for a finite lattice, by enumerating all possible configurations of the dimer problem. For even quite small lattices, however, this is a computationally difficult problem, and we have made no attempt to find $\T$ explicitly. Instead, the continuum theory for the transition can be found by using general principles such as symmetry to place constraints on $\quH$.

The Hamiltonian will take the form of a generic lattice-boson model, with all hopping terms and interactions that are consistent with symmetry, as in the related mapping for spin ice \cite{SpinIce}. The parameters in the Hamiltonian are functions of those in the classical model, and so in our case the hopping and interaction strengths will depend on the classical temperature $T$. Since the high-temperature phase corresponds to a boson condensate, we expect the ratio of the hopping strength to the interactions to increase with $T$.


For the mapping to bosons to be useful in the thermodynamic limit, we require that the Hamiltonian should be local, at least in its action on low-energy states. We have no general proof that this is the case, but we argue that it is so in the region of interest, near the transition. In this case, the low energy states of the bosons can be written in terms of states with a finite density of domain walls separating regions with different density-wave orderings. These one-dimensional domain walls are, in 3D, the intersections of a given $(111)$ plane with two-dimensional domain walls between dimer orderings. Consider a domain wall that moves by a large amount in a single time step, so that one of the two neighbouring domains grows by an area $\delta A$. In the three-dimensional picture, this domain wall has a section of area $\delta A$ that runs parallel to the $(111)$ plane. Such a configuration costs entropy and configuration energy that grows with $\delta A$, and hence has an exponentially suppressed contribution to the transfer matrix.


We defer a full analysis of the symmetry properties of the Hamiltonian $\Ham$ to a forthcoming paper, and discuss here only a symmetry of particular importance: rotation by $\frac{\pi}{3}$ about a $[111]$ axis passing through the centre of a kagome hexagon, and hence through a cubic lattice site with $\tau \bmod 3 = 0$ (such as the one at the centre of Figure~\ref{Cubic111Projection}). This maps the kagome lattice onto itself, but is not a symmetry of the cubic lattice, since it exchanges the positions of the cubic sites on the intermediate planes. (For instance, it exchanges cubic sites with $\tau = 1$ and $\tau = 2$.) We therefore define the operation $\Rot$ consisting of this rotation followed by a reflection in the $(111)$ plane $\tau = \frac{3}{2}$, which is a symmetry of the cubic lattice. In terms of the bosons, $\Rot$ is a rotation followed by a time-reversal operation $\tau \rightarrow 3 - \tau$, and it commutes with $\quH$.


Before addressing the transition, we note that taking full account of the symmetries allows an alternative derivation \cite{SpinIce}, starting from the quantum model, of the dipolar form of the Coulomb-phase correlation functions. This phase corresponds to the superfluid, where long-range correlation functions are dominated by fluctuations of the Goldstone phase mode $\phi$, described by the coarse-grained action $\Lag _\phi \sim |\partial \phi|^2$. Correlation functions can be found by using the symmetries to express the boson number operators and dimer occupations in terms of the continuum field $\phi$, giving for instance $d _\mu \sim \eta _\rv \partial _\mu \phi$.


The continuum theory describing the transition between a superfluid and a Mott insulator at fractional fillings can be expressed in terms of dual vortex fields. We will sketch the derivation of this theory; readers are referred to the papers of Balents et al.\ \cite{Balents} for an overview of the approach. In essence, it consists of a transformation from a current-loop representation of the boson problem \cite{Wallin}, where the basic degrees of freedom are the currents $J_\mu(\rv)$ defined on the links of the space-time lattice, to a gauge field $A_\mu(\rv)$ defined on the links of the dual lattice, according to $\boldsymbol{J} = \del \times \boldsymbol{A}$.

Note that the bosons belong on the sites of the \kagome\ lattice, and so the space-time lattice is not the original cubic lattice of the dimer problem, but instead consists of stacked \kagome\ planes. The dual of \kagome\ is the dice lattice \cite{Sengupta,Jiang}, and so the gauge field $\Av$ is defined on the links of a lattice of stacked dice planes.

The dual theory has a gauge invariance resulting from the redundancy of the definition of $\Av$, and can be written in terms of the gauge field $A_\mu(\rv)$ and matter fields $\psi(\rv)$ defined on the sites of the dual lattice. In terms of the original bosons, these matter fields represent vortices, and the physical density of bosons $f = \frac{1}{6}$ becomes an effective magnetic field $\del\times\boldsymbol{\bar{A}}$ for the vortices. The low-energy properties of the theory are then determined by the vortex dynamics in this effective magnetic field.

For bosons at filling $f = \frac{1}{6}$ on the kagome lattice, the conclusion of this analysis is that there are two degenerate minima of the vortex dispersion within the reduced Brillouin zone \cite{Jiang}. The continuum theory can then be written in terms of field operators $\varphi _0$ and $\varphi _1$ corresponding to these low-energy modes, and the symmetries again strongly constrain the possible terms in the action.

An important example is the symmetry under $\Rot$, which transforms the fields according to
\beq{RotationTransformation}
\begin{pmatrix}
\varphi _0\\
\varphi _1
\end{pmatrix}
\xrightarrow{\Rot}
\frac{\ee^{\ii \frac{\pi}{12}}}{\sqrt{2}}
\begin{pmatrix}
1&1\\
-\ii&\ii
\end{pmatrix}
\begin{pmatrix}
\varphi _0\\
\varphi _1
\end{pmatrix}\punc{,}
\eeq
implying that this must be a symmetry of the continuum theory. While the full action is only symmetric under a discrete subgroup of $\mathrm{SU}(2)$, the symmetry under $\Rot$ forbids any terms that break $\mathrm{SU}(2)$ up to (and including) fourth order in the fields $\phiv$.

The allowed interaction terms can be found by expressing all gauge-invariant bilinears of the fields $\varphi _0$ and $\varphi _1$ in terms of the boson density operators \cite{Balents}, or equivalently, in terms of the dimer order parameters $m _\mu$. Using the symmetries of the cubic lattice, we find
\beq{OrderParameterFromVortices}
m_\mu \sim \phiv^\dagger \sigv^\mu \phiv\nd\punc{,}
\eeq
where $\sigv^\mu$ are the Pauli matrices.

Defining the $\mathrm{SU}(2)$ Casimir invariant $\Omega = |\varphi _0|^2 + |\varphi _1|^2$, all gauge-invariant combinations of the fields $\phiv$ can be written in terms of $\Omega$ and $m _{x,y,z}$, and it is straightforward to show that $\Omega^2 \sim |\mv|^2$. While the action can contain any term involving only $\Omega$, terms involving functions of $m_\mu$ that break $\mathrm{SU}(2)$ are strongly restricted by symmetry. The first such term allowed by \refeq{RotationTransformation}, of sixth order in $\phiv$, is $m_x m_y m_z$, and is excluded by requiring symmetry under $\mv \rightarrow -\mv$. The lowest-order combination satisfying all the symmetries is of eighth order, and is given by $\Lag _1 = v\sum _\mu m_\mu^4$. In order to describe the transition to the particular ordered states with which we are concerned, we require $v < 0$.

The $\mathrm{SU}(2)$-invariant part of the action, given in \refeq{Lag0}, takes the standard form for a complex vector minimally coupled to a noncompact $\mathrm{U}(1)$ gauge field. We have omitted terms containing higher derivatives or higher powers of the field $\phiv$. The cubic symmetry of the original dimer problem means $\Lag _0$ must be space--time symmetric, and we have made this explicit in \refeq{Lag0}.

It is not firmly established whether this action has any nontrivial fixed points under the RG. As noted in Ref.~\onlinecite{Balents}, a conjecture of this sort is difficult to test analytically \footnote{For instance, an expansion in $\epsilon = 4 - d$, where $d$ is the spatial dimension, has no stable weak-coupling fixed points.}, and the evidence from numerical studies is still inconclusive. It is clear from simulations, however, that the transition is, at most, weakly first order, with a very large correlation length, and so a continuum description is appropriate. The fact that $\Lag _1$ is of eighth order in the field $\phiv$ makes it highly likely that this term is irrelevant in the continuum, and that the effective theory has an emergent $\mathrm{SU}(2) \cong \mathrm{O}(3)$ symmetry.

Using this assumption, and the expression for the crystalline order parameter given in \refeq{OrderParameterFromVortices}, we are able to explain two qualitative observations made by Misguich et al.~\cite{Misguich} based on their numerical results near the transition. Firstly, the data strongly suggest that the order-parameter distribution is spherical, as expected for an $\mathrm{O}(3)$-symmetric theory. Secondly, the dimer--dimer correlations are dominated by a `spin-like' contribution \cite{Misguich}. This follows from the fact that $m _\mu$ (and hence the dimer variables $d _\mu$) couples directly (without derivatives) to a bilinear in the critical field $\phiv$, and forms a three-dimensional representation of $\mathrm{SU}(2)$. We therefore find $\langle m_\mu m_\nu \rangle \sim \delta_{\mu\nu} |\rv|^{-d + 2 -\eta _m}$ \cite{Misguich}. The absence of a weak-coupling fixed point again prevents us from making quantitative predictions about the anomalous dimension $\eta _m$. As noted by Misguich et al.~\cite{Misguich}, these properties, while explained straightforwardly by the NC$CP^1$ theory, are incompatible with other, more obvious, candidate continuum theories, such as the $\mathrm{O}(3)$ model.

In conclusion, we have derived an NC$CP^1$ theory to describe the observed phase transition in a classical dimer model on the cubic lattice \cite{Alet}, by means of a mapping to a quantum transition in two dimensions. Simple qualitative predictions based on this mapping are in agreement with numerical results for the dimer system \cite{Alet,Misguich}. The discrepancies between the observed properties of the NC$CP^1$ and dimer models, as well as the uncertainty regarding the nature of the transitions, illustrate the need for further numerical studies of both models.

We thank P.\ Fendley and L.\ Balents for helpful comments. This work was supported in part by EPSRC Grant No.\ EP/D050952/1.

\end{document}